\documentclass[twocolumn,amsmath,amssymb,nofootinbib,tighten,floatfix,prb]{revtex4}

\usepackage{color}
\usepackage{graphicx}
\usepackage{dcolumn} 
\usepackage{bm, bbm}      
\usepackage{curves}
\usepackage{epic}
\usepackage{wasysym}
\usepackage{epsf, times}
\usepackage{subfigure}
\usepackage{eufrak}
\usepackage{amsthm}
\usepackage{ulem} 

\begin{document}

\title{Isolated Flat Bands and Spin-1 Conical Bands in Two-Dimensional
Lattices}

\author{Dmitry Green$^{1}$}
\email{dmitrygreen2009@gmail.com}
\author{Luiz Santos $^{2}$}
\author{Claudio Chamon$^{3}$}

\affiliation{$^{1}$\!\!\!
BlueMountain Capital Management LLC,
New York, New York 10017, USA
\\
$^{2}$\!\!\!
Department of Physics, Harvard University,
17 Oxford Street,
Cambridge, Massachusetts 02138, USA
\\
$^{3}$\!\!\!
Physics Department, Boston University, 590 Commonwealth
Avenue, Boston, Massachusetts 02215, USA
}

\date{\today}

\begin{abstract}
Dispersionless bands, such as Landau levels, serve as a good 
starting point for obtaining interesting correlated states
when interactions are added. With this motivation in mind,
we study a variety of dispersionless (``flat'') band structures that
arise in tight-binding
Hamiltonians defined on hexagonal and kagome lattices with staggered
fluxes. The flat bands and their neighboring dispersing bands have
several notable features: (a) flat bands can be isolated from other
bands by breaking time-reversal symmetry, allowing for an
extensive degeneracy when these bands are partially filled; (b) an
isolated flat band corresponds to a critical point between regimes
where the band is electron-like or hole-like, with an anomalous Hall
conductance that changes sign across the transition; (c) when the gap
between a flat band and two neighboring bands closes, the system is
described by a single spin-1 conical-like spectrum, extending to higher angular
momentum the spin-1/2 Dirac-like spectra in topological insulators and
graphene; (d) some configurations of parameters admit two isolated parallel flat bands, raising the possibility of exotic ``heavy excitons'' and (e) we find that the Chern number of the flat bands, 
in all instances that we study here, is zero.

\end{abstract}

\maketitle

\def\openone{\leavevmode\hbox{\small1\kern-4.2pt\normalsize1}}

\newcommand{\evec}{\mathbf e}
\newcommand{\kvec}{\mathbf k}
\newcommand{\pvec}{\mathbf p}
\newcommand{\svec}{\mathbf s}
\newcommand{\Rvec}{\mathbf R}
\newcommand{\rvec}{\mathbf r}
\newcommand{\bsigma}{\mathbf \sigma}
\newcommand{\ii}{{\rm i}}

\newcommand{\slapar}{\not \hskip -2 true pt \partial\hskip 2 true pt}
\newcommand{\slapartxt}{\not\!\!\!\!\partial}
\newcommand{\slaA}{\!\not\!\! A}
\newcommand{\slaAA}{\!\not\!\! A^{5}}
\newcommand{\slaM}{\ \backslash \hskip -8 true pt M}
\newcommand{\slaS}{\ \backslash \hskip -7 true pt \Sigma}

\newcommand{\au}{\alpha_+}
\newcommand{\ad}{\alpha_-}
\newcommand{\aub}{{\overline \alpha_+}}
\newcommand{\adb}{{\overline \alpha_-}}



%
%

\section{Introduction}
\label{sec:intro}

One of the reasons why dispersionless (or flat) bands are
interesting is that they accommodate, when partially filled, an
exponentially large number of states. This macroscopic degeneracy can
be lifted when interactions are added, often leading to rich strongly
correlated phenomena. The best known example is the fractional
quantum Hall effect, which arises from the degeneracy within flat
Landau bands for particles in a magnetic field.

In addition to the Landau problem, other models with flat bands have
been studied at least since the 1970s, such as amorphous
semiconductors~\cite{Weaire1,Weaire2,Thorpe}. This system is idealized
by a lattice made up of clusters of sites.  Both inter-cluster and
intra-cluster hoppings are allowed. From a mathematical point of view,
it turns out that the intra-cluster hopping term in the Hamiltonian is
a projection operator, leading to the existence of flat
bands~\cite{Straley,Straley_footnote}.

In the 1980s, flat bands were studied in relation to the
Nielson-Ninomiya theorem~\cite{Nielson} by Dagotto \textit{et al}~\cite{Dagotto}.  They showed that it is possible to escape the
fermion-doubling problem at the price of having an extra flat band in
the spectrum. In this way, the low energy degrees of freedom of the
theory can be described as a single Weyl species.

More recently, Ohgushi \textit{et al}~\cite{Nagaosa} studied
flux phases in the kagome lattice, which is the planar section of
ferromagnetic textured pyrochlores. If the flux is staggered then
electrons accumulate a spin Berry phase as they hop. This system
contains {\it isolated} flat bands, {\it i.e.}, they are protected by a gap.
On the other hand, Bergman \textit{et al} have studied flat bands
{\it without a gap} in similar lattices~\cite{Wu2} not threaded by
fluxes. In their models, a flat band is degenerate with one or more
other bands at a single point, and the touching is topologically
protected. Each of the works above has identified interesting, but
seemingly disconnected, properties of flat bands.

The purpose of this paper is to understand the different types of flat
band spectra, the conditions to obtain them, and the properties that
follow. This paper is organized around five main findings.  First,
flat bands can be isolated by breaking time-reversal symmetry (TRS).  
Second, isolated flat bands can be viewed as critical points. On
either side of the phase boundary the flat band becomes positively or
negatively curved, which corresponds to transitioning from a
particle-like to a hole-like band. We also find an anomalous Hall
effect on either side of the transition, whose sign depends on whether
the band is electron-like or hole-like. Third, when the gap between
a flat band and its neighboring bands closes with the flat band in the
middle, the system is described by a single spin-1 conical-like
spectrum. This extends to higher angular momentum the spin-1/2
Dirac-like spectra in topological insulators and in graphene.
Fourth, one can obtain multiple (parallel) flat bands and we provide 
a concrete example of this case. Fifth, we find, for all examples
studied here, that the Chern number of the flat bands is zero.
So as opposed to Landau levels, we get no quantized Hall 
conductance for these particular flat bands.

\begin{figure}
(a)\includegraphics[angle=0,scale=0.3]{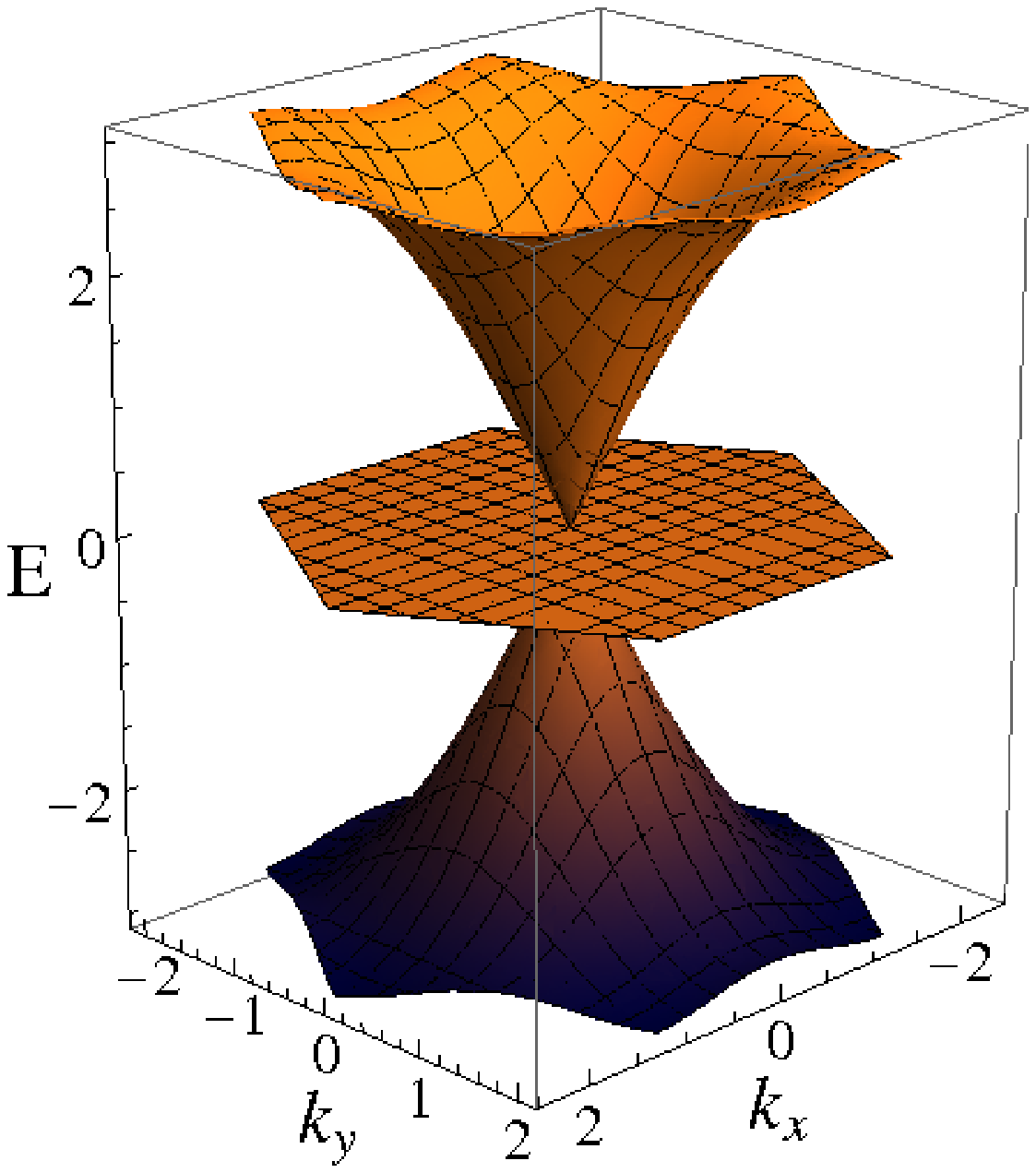}
(b)\includegraphics[angle=0,scale=0.3]{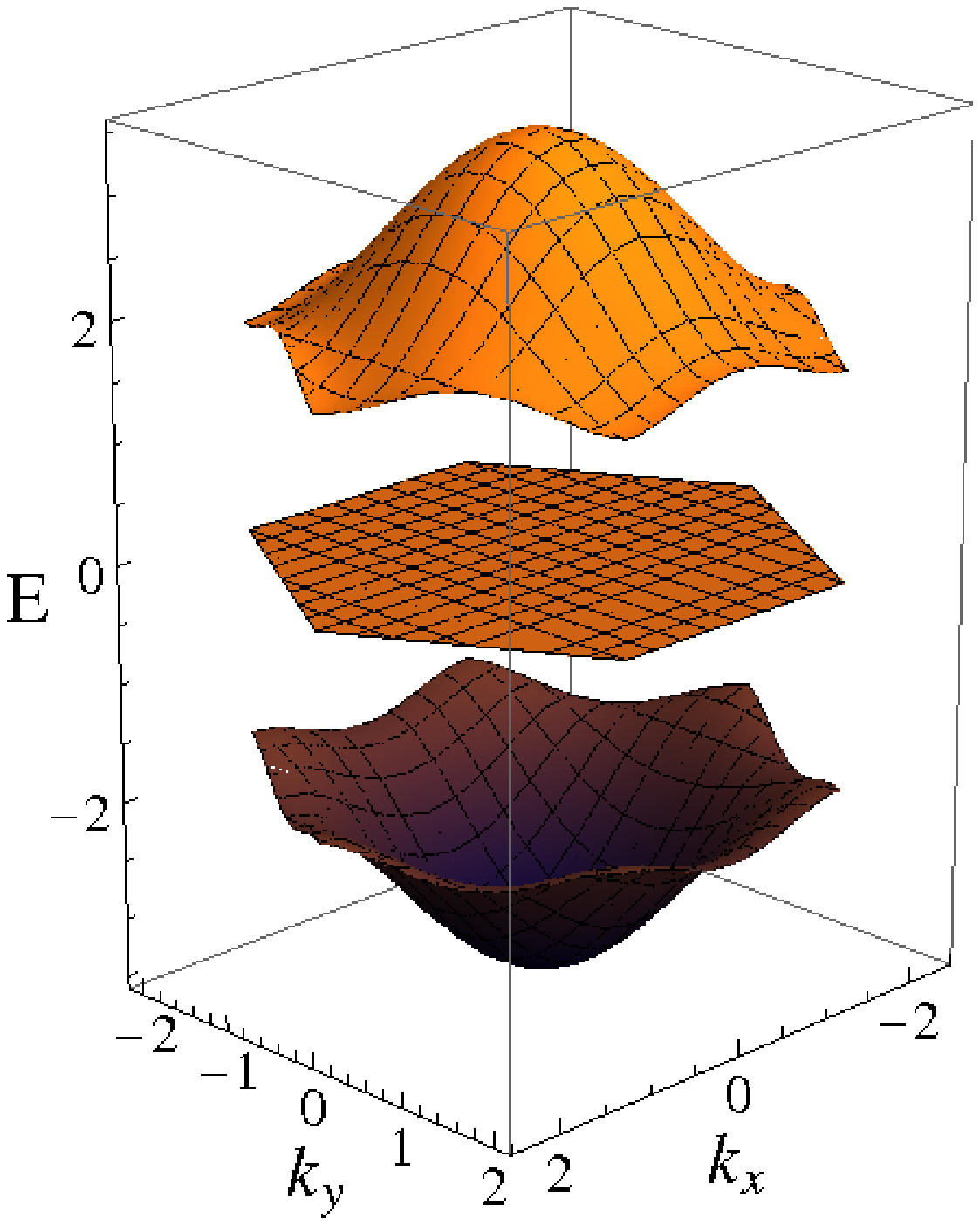}
\caption{ Energy dispersions with
staggered flux phases $\phi_+$ and $\phi_-$ on the up and down triangles of the kagome lattice . The dispersion on the left (type I) corresponds
to $\phi_+ = 2\pi$ and $\phi_-=-\pi$ while the dispersion on the right (type II) corresponds to $\phi_+ = \phi_- = 3\pi/2$.
         }
\label{fig:kagome-zero}
\end{figure}

The model with which we will start our analysis is a simple kagome
lattice with a tight-binding interaction and two staggered fluxes,
$\phi_+$ and $\phi_-$, on alternating triangles. 
Depending on the values of $\phi_\pm$, dispersions can be
classified into three types:
(I) a flat band touches two
linearly dispersing bands at the same point, where the linear bands
are reminiscent of a ``Dirac-like" point, but with spin-1 behavior,
(II) an isolated flat band that is separated
from bands above and below by a gap, and
(III) a gapless flat band that touches a single massive energy band either
above or below. The energy dispersions corresponding to types I and II are plotted in Fig. \ref{fig:kagome-zero}.
Type III   has been discussed recently~\cite{Huse,Wu1,Wu2} in
hexagonal and kagome lattices without magnetic flux, and it was found
that in this case the zero gap is protected by topological
arguments~\cite{Wu2}. Type III has been also
shown to exhibit a topological insulator phase in the presence
of spin-orbit interactions~\cite{Guo2009}. Type II appears when electrons
accumulate a spin Berry phase as they hop from site to
site, which is equivalent to both up and down triangles
with the same magnetic flux~\cite{Nagaosa}. Type I  necessitates the
staggered fluxes $\phi_+\ne\phi_-$ that we analyze below.

We will show that the condition for a flat band to occur at $E=0$ is
$\phi_++\phi_-\equiv \pi\, ({\rm mod}\;2\pi)$. By changing the value
of the fluxes in such a way that their sum differs slightly from
$\pi$, the flat band acquires a small curvature, which can be positive
or negative depending on the values of $\phi_\pm$. Interestingly, the
band curvature implies that if the Fermi energy is chosen to be zero,
then by tuning the fluxes it is possible to change the center of the
band from an electron-like pocket to a hole-like pocket.  This leads
to an inversion of the sign of the anomalous Hall
response~\cite{Haldane2004}.  Therefore, the flat band condition
$\phi_++\phi_-\equiv \pi\, ({\rm mod}\;2\pi)$ represents a quantum
critical point separating two regions with different anomalous Hall responses.

Type I is remarkable in that it displays linearly dispersing modes,
akin to those of graphene, but differing in two important ways. First,
the conical points do not appear in pairs as in graphene, but instead
there is only one such point within the first Brillouin zone (BZ) (see
Fig. \ref{fig:kagome-zero}). Second, these are not Dirac fermions
(this is why it is possible to evade the doubling problem), but
instead the effective Hamiltonian in momentum space is of the form
$H=v_F\,{\vec k}\cdot{\vec L}$, where $\vec L$ is the spin-1 angular
momentum operator ($v_F$ is the Fermi velocity). The spin-1-type
spectrum (like the spin-1/2 Dirac-type spectrum of graphene) can be
viewed as a single quantum spin in a magnetic field, as in Berry's
original work on quantum phases~\cite{Berry}, but with the wave vector
${\bf k}$ playing the role of the magnetic field. Type I does not
require the breaking of TRS. However, we show that a gap can be opened
while leaving the flat band untouched by breaking TRS, and thus type
I is continuously connected to type II . In type II the degenerate
states within the flat band are protected by the gap at finite
temperature, and would provide a fertile base to construct correlated
states.

Finally, we will generalize the above results on the kagome lattice to
other lattices by a formulation similar to Straley~\cite{Straley}, but
again adding staggered fluxes. As an example, we take a hexagonal
network, similar to graphene, but with three degrees of freedom at
each site.  We will consider two hopping strengths in this case, one
for inter-site hopping between nearest neighbors ($t$) and one on each
vertex for intra-site permutations between the three species ($g$).
Each permutation will be associated with a flux $\phi_\pm$ on the two
sublattices of the honeycomb lattice.  We will show that the kagome
model can be obtained from this in the limit $t\gg g$.  Additionally,
we will show that the honeycomb lattice admits {\it two} parallel flat
bands that are isolated from each other and all other bands, which is
a feature that the kagome lattice does not have.

\section{Flat zero-mode band in the staggered-flux kagome lattice}
\label{sec:kagome}
\begin{figure}
\includegraphics[angle=0,scale=0.4]{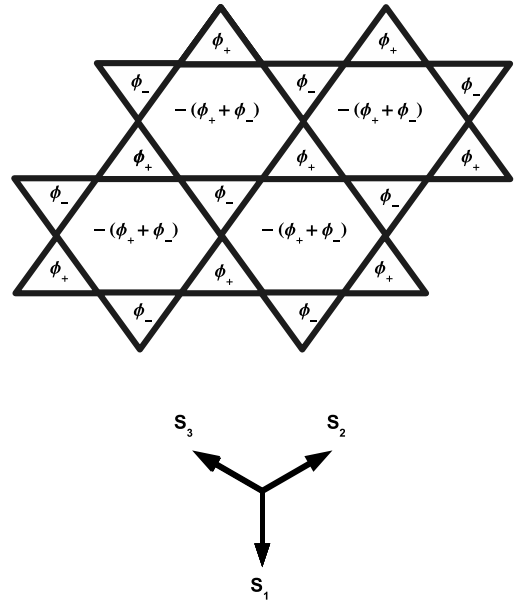}
\caption{The kagome lattice with fluxes $\phi_+$ and $\phi_-$ on alternating triangles,
which correspond to a flux $-(\phi_+ + \phi_-)$ inside each hexagon. $\svec_{1,2,3}$
are vectors pointing from the center
of a down triangle to its three up neighbors.}
\label{fig:kagome_lattice}
\end{figure}

Consider the tight-binding Hamiltonian defined on a kagome lattice,
where staggered fluxes $\phi_+$ and $\phi_-$ are applied within alternating triangles (``up'' and
``down'' triangles, respectively), as shown in Fig. \ref{fig:kagome_lattice}.

For convenience define the phase factors
$\alpha_\pm=e^{i\phi_\pm/3}$.  Let $\svec_1 = (0,-1)$, $\svec_2 = (\sqrt{3}/2,1/2)$, and
$\svec_3 = (-\sqrt{3}/2,1/2)$ be the vectors
pointing from the centers of an up triangle to its three down
neighbors, and define $d^{\,ij}_\kvec=e^{-i\kvec\cdot(\svec_i-\svec_j)}$, with $j=1,2,3$. In momentum space, the
Hamiltonian can be written as:
\begin{eqnarray}
H_{\kvec}=g \left(\begin{array}{ccc}
0&\au + \ad d^{12}_\kvec & \aub + \adb d^{13}_\kvec\\
\aub + \adb d^{21}_\kvec&0&\au + \ad d^{23}_\kvec\\
\au + \ad d^{31}_\kvec&\aub + \adb d^{32}_\kvec&0
\end{array}\right)
\;,
\end{eqnarray}
where $g$ is the hopping strength.  The characteristic polynomial for
this matrix is
\begin{equation}
\label{eq:charac_polynomial}
P(E)=-E^3+g^2 a_1(\kvec)\,E+g^3 a_0(\kvec)
\;,
\end{equation}
where
\begin{eqnarray*}
a_1(\kvec)&=&
[3+\aub\ad\,q(\kvec)]+c.c.
\,
\\
a_0(\kvec)&=&
[(\au^3+\ad^3)+(\au^2\ad+\aub\adb^2)\,q(\kvec)]+c.c.
\;,
\end{eqnarray*}
and $q(\kvec)=d^{12}_\kvec +d^{23}_\kvec +d^{31}_\kvec $. The
condition for a flat ($\kvec$-independent) band is obtained by setting
the overall factor of $q(\kvec)$ in $P(E)$ to zero,
\begin{equation}
\label{eq:no_k_dependence}
g^2\aub\ad E +g^3(\au^2\ad+\aub\adb^2)=0
\;, 
\end{equation}
which is equivalent to 
\begin{equation}
\label{eq:flatband_energy}
E=-g(\au^3+\adb^3)
\;.
\end{equation}
Combining Eqs.~(\ref{eq:charac_polynomial})-(\ref{eq:flatband_energy}) gives the following equation for the
energy eigenvalues of the flat bands:
\begin{equation}
\label{eq:flatband_energy2}
E(E-4g^2)=0
\;,
\end{equation}
which has three possible solutions:\\

\textbf{(a)} $E=0$ flat band: this configuration is achieved for
$\phi_+ + \phi_- =\pi\, ({\rm
mod}\;2\pi)$,\\

\textbf{(b)} $E = -2g$ flat band: this configurations is achieved 
for $\phi_{\pm} = 2\pi n_{\pm}$, with $n_{\pm}$ integer
valued numbers, and\\

\textbf{(c)} $E = 2g$ flat band: this configurations is achieved 
for $\phi_{\pm} = \pi(2 n_{\pm}+1)$, with $n_{\pm}$ integer
valued numbers.\\ 

The cases $E=\pm 2g$ (type III) are similar to
those discussed in Refs.~\cite{Huse,Wu1,Wu2}, where the flat band
touches a parabolic electron-like band at its bottom (for $E=-2|g|$)
or a hole-like band at its top (for $E=+2|g|$). Here, we shall focus
instead in the case where the flat band is at $E=0$.
In Fig.~\ref{fig:kagome-zero} (types I and II) we show two
particular choices for $\phi_\pm$, which are representative of what we
classify as types I and II spectra. Type I contains a cone vertex
touching at the point $\kvec=0$, which we illustrate by setting
$\phi_+=3\pi$ and $\phi_-=0$ (this choice of phase can be interpreted
as tight-binding hoppings $-g$ for up triangles and $+g$ for down
triangles). Type II contains an isolated flat band, which we
illustrate by setting $\phi_\pm=3\pi/2$ (this choice can be
interpreted as tight-binding matrix elements $\pm ig$ for hopping
anti-clockwise or clock-wise around the triangles). Notice that
because ${\rm tr} H_{\kvec}=0$ and one of the eigenvalues is $E=0$,
the other two eigenenergies must satisfy $E_+(\kvec)+E_-(\kvec)=0$, so
the spectrum is symmetric with respect to zero in both types I and
II.

The general condition for nodal touching (type I) can be obtained by
requiring that there is another $E=0$ eigenvalue, so that at least one
other band touches the flat band. When such a solution exists, the
derivative of the characteristic polynomial $P'(E)$ also has a zero at
$E=0$ for some value of $\kvec$. This condition translates to
\begin{equation}
a_1(\kvec)=0 \Rightarrow \aub\ad q(\kvec)=-3
\;,
\end{equation}
which admits three different solutions:\\

\textbf{(A)} Nodal point at $\Gamma = (0,0)$: 
this type I configuration is obtained if the condition
$\phi_+ - \phi_- = 3\pi + 6\pi n$ is satisfied, where $n$ is an integer,
and it is illustrated in Fig. \ref{fig:kagome-zero} (type I),\\

\textbf{(B)} Nodal point at $\kvec = K_{+} = (\frac{4\pi}{3\sqrt{3}},0)$: 
this type I configuration is obtained if the condition
$\phi_+ - \phi_- = 5\pi + 6\pi n$ is satisfied, where $n$ is an integer, and\\

\textbf{(C)} Nodal point at $\kvec = K_{-} = (-\frac{4\pi}{3\sqrt{3}},0)$: 
this type I configuration is obtained if the condition
$\phi_+ - \phi_- = \pi + 6\pi n$ is satisfied, where $n$ is an integer.\\

Even though types I and II are particle-hole symmetric with a flat band
at $E=0$ and have staggered fluxes obeying the constraint 
$\phi_+ + \phi_- =\pi\, ({\rm mod}\;2\pi)$, type II lacks the
nodal conditions \textbf{(A)-(C)} mentioned above. 
 
\subsection{Nodal touching and the spin-1 cone}

Let us now expand the Hamiltonian, in type I , near the vertex point
for small $|\kvec|=\sqrt{k_x^2+k_y^2}$.  At the same time we move into
type II by applying a slight flux offset from the condition for the
touching: $\phi_+=3(\pi+\delta)$ and $\phi_-=-3\delta$.  We will
interpret $\delta$ as a ``mass'' term.  To first order in $\delta$ and
$\kvec$ the Hamiltonian becomes:
\begin{eqnarray}
H_{\kvec}&=&g\, \frac{3}{\sqrt{2}}\;
\left[
k_x\;L'_x+
k_y\;L'_y+
2\sqrt{\frac{2}{3}}\; \delta\;L'_z
\right]
\nonumber\\
&=&g\, \frac{3}{\sqrt{2}}\; \left(k_x,k_y,m\right)
\cdot \vec L'
\;,
\label{eq:p-dot-L}
\end{eqnarray}
where $m=2\sqrt{{2}/{3}}\; \delta$. It is straightforward to check that
the matrices

\begin{subequations}
\label{eq:spin-1 matrices}
\begin{equation}
L'_x=
\frac{i}{\sqrt{6}}\left(
\begin{array}{ccc}
 0 & 1& -1 \\
 -1 & 0 & -2 \\
1& 2& 0
\end{array}
\right)
\end{equation}
\begin{equation}
L'_y=
\frac{i}{\sqrt{2}}\left(
\begin{array}{ccc}
 0 & 1 & 1 \\
 -1 & 0 & 0 \\
 -1 & 0 & 0
\end{array}
\right)
\end{equation}
\begin{equation}
L'_z=\frac{i}{\sqrt{3}}\left(
\begin{array}{ccc}
 0 & -1 & 1 \\
 1 & 0 & -1 \\
 -1 & 1 & 0
\end{array}
\right)
\end{equation}
\end{subequations}
satisfy the angular momenta algebra $[L'_x,L'_y]=iL'_z$ (along with
the cyclic permutations of $x$, $y$, and $z$), and that they have
eigenvalues $-1,0,+1$, {\it i.e.}, they form a spin-1 representation
of SU($2$).

The eigenvalues of the Hamiltonian (\ref{eq:p-dot-L}) are
$E_{\kvec}=g\,\frac{3}{\sqrt{2}}\;\sqrt{k_x^2+k_y^2+m^2}\;\;\ell_\kvec$,
where $\ell_\kvec=-1,0,+1$ is the eigenvalue of angular momentum along
the direction $\left(k_x,k_y,m\right)$.  Therefore we obtain the three
bands, with the flat band being the one with zero angular
momentum. The other two bands describe the cone when $\delta=0$, and
two parabolic bands separated from the flat band by a gap
$\Delta=g\,2\sqrt{3}\,\delta$ when $\delta$ is non-zero. (A particular instance of the $\delta = 0$ spectrum has recently also been predicted in ${\cal T}_3$ optical lattices by Bercioux \textit{et al.}~\cite{Bercioux})

The spin-1 structure has interesting topological properties, namely it
can be viewed as a generalization of the Berry phase for the spin-1/2
spectrum of graphene. One implication is that, when the gap is open by
breaking TRS, the upper and lower bands have a quantized Hall
conductance. In the Appendix we explicitly compute the
Chern number over the first Brillouin zone for the three bands.

\subsection{Transitioning between electron and hole bands}

\begin{figure}
(a)\includegraphics[angle=0,scale=0.3]{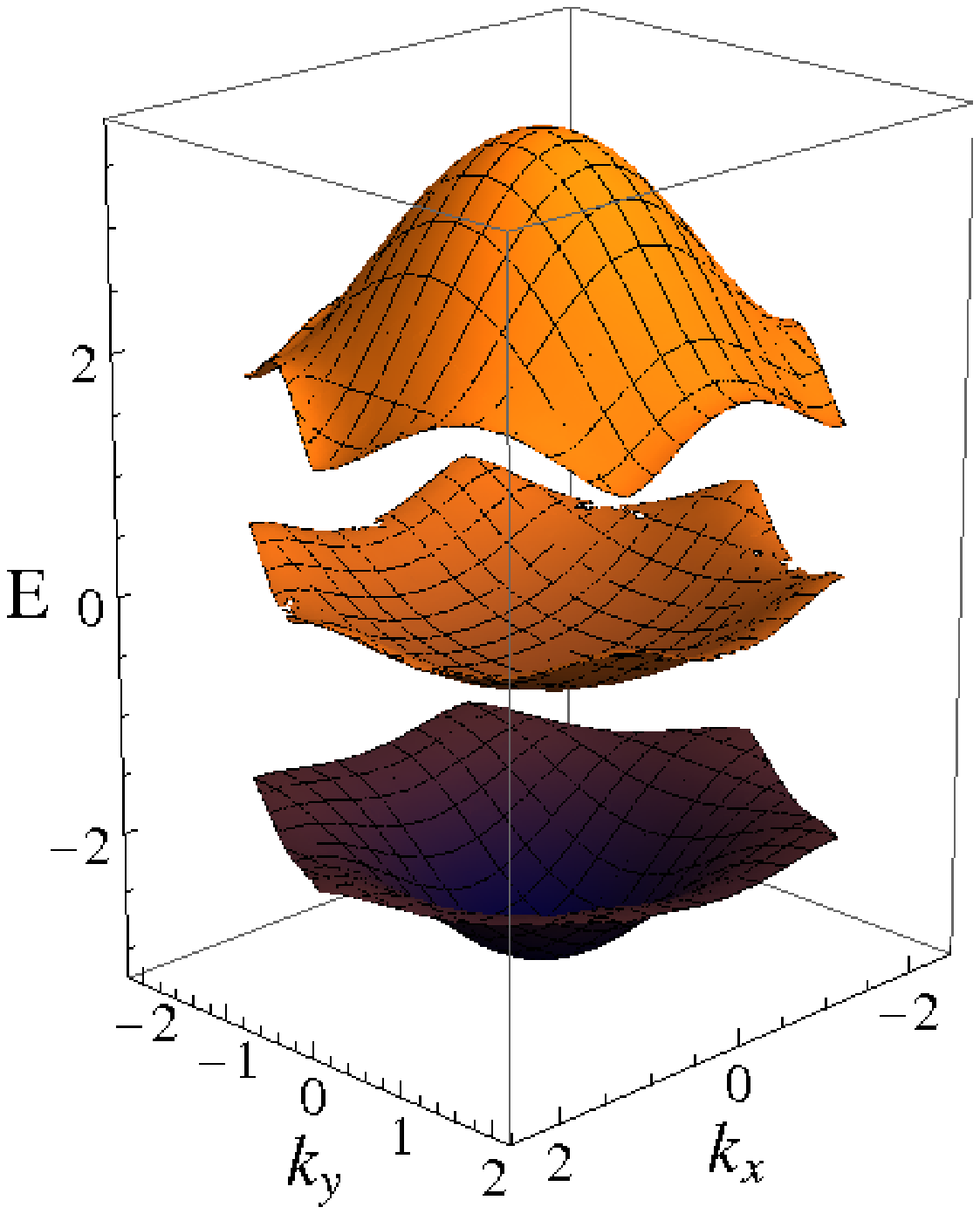}
(b)\includegraphics[angle=0,scale=0.3]{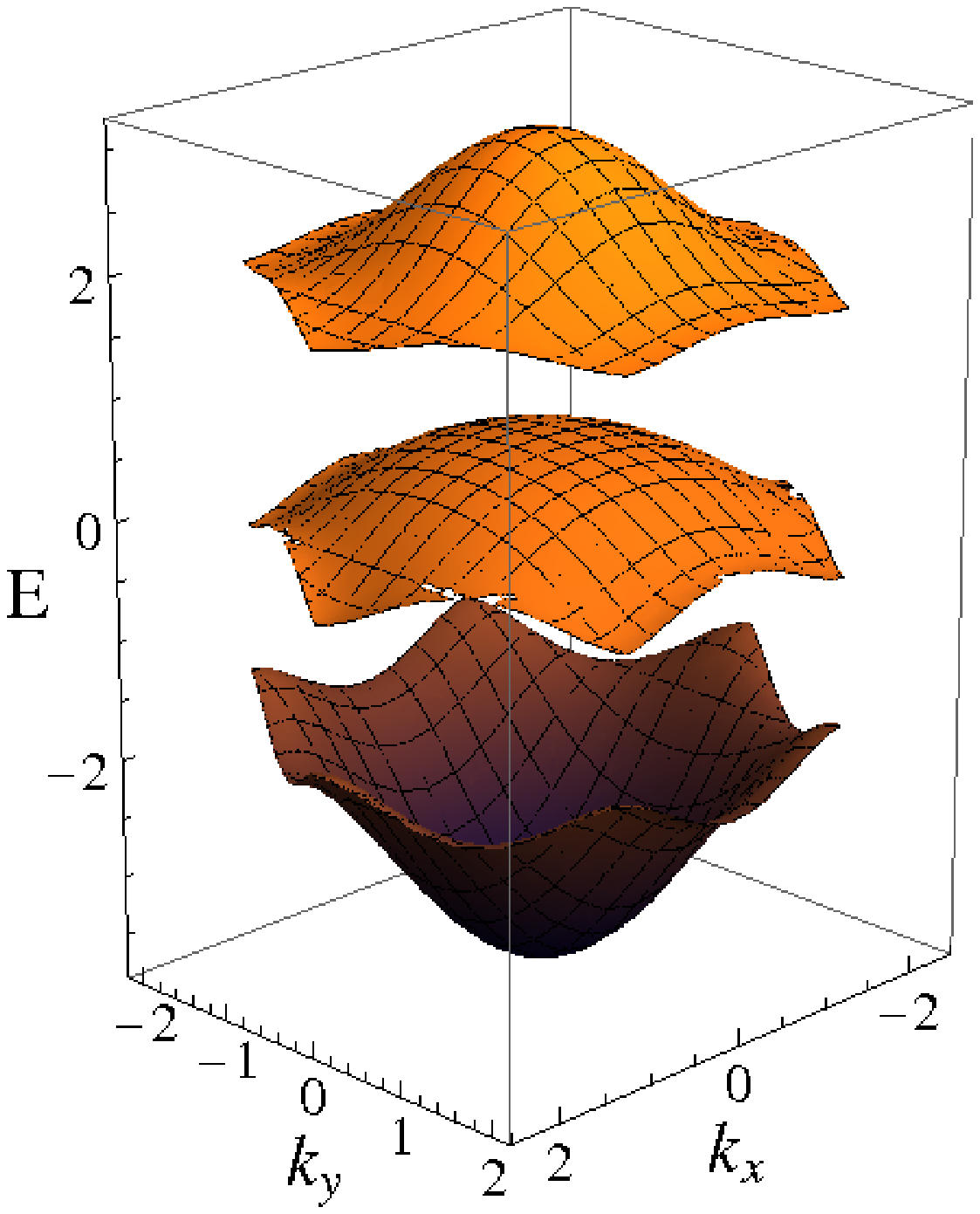}
\caption{Left (right): middle band showing electron-like (hole-like)
dispersion corresponding to $\epsilon < 0 (> 0)$ in
Eq.~(\ref{eq:band_curvature}).}
\label{fig:electron-hole dispersion}
\end{figure}

We now consider deviations from the flat band condition for the case
when the middle band is isolated, as in type II . For concreteness,
consider the case $\phi_+=\phi_-=3(\pi/2-\epsilon)$. For small
$\epsilon$, the energy of the middle band will be close to $E=0$, so we
can obtain the dispersion for the middle band by dropping the cubic
term in the characteristic polynomial $P(E)$,
\begin{equation}
E(\kvec)= -g\,\frac{a_0(\kvec)}{a_1(\kvec)}+{\cal O}(\epsilon^3)
\;.
\end{equation}
Expanding $a_0(\kvec),a_1(\kvec)$ up to order $|\kvec|^2$ and
$\epsilon$, one obtains
\begin{equation}
\label{eq:band_curvature}
E(\kvec)= g\,\left(4\epsilon-\frac{3\epsilon}{4}\; |\kvec|^2\right)
\;,
\end{equation}
corresponding to a band mass $m_\epsilon=-3\epsilon/2\,g$ for the
middle band, so it has a hole-like dispersion for $\epsilon>0$ and
an electron-like dispersion for $\epsilon<0$, as depicted in
Fig. \ref{fig:electron-hole dispersion}. This trivial
mathematical result is physically remarkable in that one can, in
principle, change the character of a band from electron-like to
hole-like by varying one parameter in the Hamiltonian.

Notice that there is also a band shift $4\epsilon g$, which adds to the
chemical potential. There is an interesting result when the chemical
potential is fixed to $\mu=0$: the Fermi surface is pinned and
independent of $\epsilon$. It is best to see this effect prior to any
perturbation in $\epsilon$ or expansion in $|\kvec|$. The Fermi
surface is in this case the locus of $\kvec$ points for which
$P(E=0)=0$, those that satisfy $a_0(\kvec)=0$. For the electron-like
case, the Fermi sea is in the region bounded by the $a_0(\kvec)=0$
surface that contains the $\Gamma$-point, whereas for the hole-like
case the Fermi sea is the complementary region in the Brillouin
zone. The anomalous Hall effect is given by the integral of the Berry
curvature over the Fermi sea. As we show in the Appendix,
the Chern number, the Berry curvature integrated over the complete
Brillouin zone, for the middle flat band is zero. This means that the
sum of the anomalous Hall effects for the electron-like and hole-like Fermi
seas is zero. Thus, as one tunes across holding $\mu=0$, the anomalous
Hall effect will change sign. Of course, by tuning $\mu$ one can vary
the anomalous Hall effect continuously.

\subsection{Time-reversal and particle-hole symmetries}

First, we discuss time-reversal symmetry.  Consider a tight-binding
model of spinless fermions described by
\begin{equation}
\mathcal{H}=\sum_{\kvec}\psi_i^{\dagger}(\kvec) H_{ij}(\kvec) \psi_j(\kvec)\;,
\end{equation}
where $\psi$ is an annihilation fermionic operator and $\kvec$ takes
values on the first Brillouin zone. Under time-reversal
transformation,
\begin{eqnarray}
\mathcal{H} & \rightarrow &
\sum_{\kvec}\psi_i^{\dagger}(-\kvec) H_{ij}^{*}(\kvec) \psi_j(-\kvec)
\nonumber\\
&=&\sum_{\kvec}\psi_i^{\dagger}(\kvec) H^{*}_{ij}(-\kvec) \psi_j(\kvec)\;.
\end{eqnarray}
For $\mathcal{H}$ to be time-reversal invariant, one way would be to
have $H(\kvec)=H^{*}(-\kvec)$. Looking at our Hamiltonian more
carefully, though, we see that there is a freedom to redefine the
hopping matrix elements without changing the fluxes $\phi_+$ and
$\phi_-$. This freedom can be mathematically described by the
following gauge transformation:

\begin{equation}
\label{eq:gauge freedom}
H(\kvec) \rightarrow \tilde{H}(\kvec)=\Lambda H(\kvec) \Lambda^{\dagger}\;,
\end{equation}
where

\begin{equation}
\Lambda= \left(\begin{array}{ccc}
e^{i\alpha_1} & 0 &0 \\
0 & e^{i\alpha_2} &0 \\
0 & 0 & e^{i\alpha_3}
\end{array}\right) \;.
\end{equation}
Hamiltonians $H(\kvec)$ and $\tilde{H}(\kvec)$ related by the gauge
transformation Eq.~(\ref{eq:gauge freedom}) represent physically
equivalent descriptions of the system.  The requirement of time-reversal
symmetry, taking into account the gauge invariance given by
Eq.~(\ref{eq:gauge freedom}), becomes then
\begin{equation}
H(\kvec)=\Lambda H^{*}(-\kvec) \Lambda^{\dagger}\;,
\end{equation}
from which we get the following conditions
\begin{equation}
\label{eq:fluxes_timereversal}
e^{(2i/3)\phi_{\pm}}=e^{i(\alpha_1-\alpha_2)}=e^{i(\alpha_2-\alpha_3)}=e^{i(\alpha_3-\alpha_1)}\;.
\end{equation}
Equation (\ref{eq:fluxes_timereversal}) immediately implies that symmetry
under time reversal is satisfied if $\phi_+=n_+\pi$ and
$\phi_-=n_-\pi$, for integers $n_+$ and $n_-$ such that $n_+ - n_- =
3l$ ($l$ integer).  This is exactly equivalent to the condition for
$\phi_{\pm}$ such that the spectrum has a gapless flat band at $E=0$,
{\it i.e.}, type I .  Therefore, we conclude,
for this given model, that in order to have an
isolated flat band \textit{time-reversal symmetry must be broken}.

Now consider particle-hole symmetry. The key observation is the
following: when the spectrum of $\mathcal{H}$ has three bands, as in
the kagome lattice, particle-hole symmetry only exists when there is a
flat band at $E=0$ and the two other bands have opposite energies. We
have already worked out the conditions for the existence of a flat
$E=0$ band in the kagome lattice to be $\phi_+ + \phi_- =\pi$, which
also dictates the conditions for particle-hole symmetry (if the
Hamiltonian is a traceless matrix). Notice that $\mathcal{H}$ can be
particle-hole symmetric without being time-reversal invariant. (A
spin-1 cone is a situation where both symmetries are present.) We
will use this important aspect later when we calculate the Chern
number of the bands.

\section{Hexagonal Lattice Model}

Having discussed the main properties of the kagome model, we now
consider an alternative model defined on the hexagonal lattice. We
will show that these models are closely related and that the kagome
model is a limiting case of the hexagonal one. Even more interestingly
the spectrum of the hexagonal model contains {\it two parallel flat
bands} that are separated from each other and from all other bands by
a gap.

\subsection{Definition of the model}

Consider a tight-binding model on a hexagonal lattice, where the
particles have three flavors. Define
the six-dimensional basis of particle operators
$(\psi^\dagger_{a,\mu}$,$\psi^\dagger_{b,\mu})$, where $\mu=1,2,3$ is
the flavor index, and $a,b$ are the two sublattices. In real space the Hamiltonian is:
\begin{eqnarray}
H&&=t\sum_{\langle ab\rangle;\mu}
\psi^\dagger_{a,\mu}
\;h_\mu(\rvec_b-\rvec_a)\;
\psi_{b,\mu}+{\rm H.c.}
\\
&&\!\!\!\!\!\!\!\!
+g\sum_{a,b;\mu,\nu}\psi^\dagger_{a,\mu}B_{\mu\nu}\alpha_+\psi_{a,\nu}
+\psi^\dagger_{b,\mu}B_{\mu\nu}\alpha_-\psi_{b,\nu}
+{\rm H.c.}
\nonumber
\label{eq:H_hex_real}
\end{eqnarray}
The first term (coupling $t$) is a nearest-neighbor hopping between
the two sublattices that conserves the flavor index, but has
correlated flavor-direction hopping controlled by $h_\mu$,
\begin{eqnarray}
h_\mu(\rvec_b-\rvec_a)=
\begin{cases}
1,& \text{${\rvec}_b-{\rvec}_a=\svec_\mu$}
\\
0,& \text{otherwise}
.
\end{cases}
\;,
\end{eqnarray}
The second term (coupling $g$) is the on-site phase dynamics, where
$B$ is the $3\times 3$ cyclical permutation matrix,
\begin{eqnarray}
B=\left(\begin{array}{ccc}
0&1&0\\
0&0&1\\
1&0&0
\end{array}\right)
\;,
\quad B^2=B^\dagger,
\quad B^3=\openone
\;,
\end{eqnarray}
and the on-site flavor changing phase factors are
$\alpha_\pm=e^{i\phi_\pm/3}$ (using the notation of Sec. II). In
other words, inter-site hopping conserves the flavor index.  Flavors
can be permuted on-site, and each permutation is accompanied by an
on-site phase factor $\phi_\pm$. This Hamiltonian can be diagonalized
analytically for particular values of $\phi_\pm$.

\subsection{Equivalence to the kagome model}

Let us write the wave function basis as
$(\psi^\dagger_{a,1},\psi^\dagger_{b,1},\psi^\dagger_{a,2},\psi^\dagger_{b,2},\psi^\dagger_{a,3},\psi^\dagger_{b,3})$
and work in momentum space.  In this basis, our Hamiltonian takes the
following six-dimensional form:

\begin{eqnarray}
H_\kvec= \left(\begin{array}{ccc}
T_1 & G & G^\dagger\\
G^\dagger & T_2 & G\\
G & G^\dagger  & T_3\\
\end{array}\right),
\label{eq:H_hex_k1}
\end{eqnarray}

where

\begin{eqnarray}
T_\mu= t\left(
\begin{array}{cc}
 0 & e^{i\kvec\cdot \svec_\mu} \\
 e^{-i\kvec\cdot \svec_\mu} & 0
\end{array}
\right)
,\;\;
G=
g\left(\begin{array}{cc}
 \alpha_+ & 0 \\
 0 & \alpha_-
\end{array}
\right)
\,
\nonumber\\
\end{eqnarray}

In the extreme case when $g=0$, $H$ is block diagonal in the
$T$ matrices and is trivial.  It has two triply degenerate eigenvalues
$E = \pm t$ and eigenvectors $v^\dagger_{\mu,\pm}(\kvec) = (1, \; \pm
e^{i\kvec\cdot \svec_\mu})/\sqrt{2}$.  This corresponds to a
``dimerized'' state since the particles cannot hop between
sublattices. (Recall that hoping between sublattices is directionally
controlled by the flavor, so if the particles are not allowed to
change flavor in the vertices because $g=0$, they can only hop back
and forth within a given bond between an $a$ and a $b$ site.)

Now project the full Hamiltonian to the lowest energy states, $E=-t$,
and expand to first order in $g/t$.  The projected Hamiltonian, $\bar
H$, is given by the matrix elements $\bar H_{\mu\nu} =
v^\dagger_{\mu,-}\,H\,v_{\nu,-}$.  For example, $\bar H_{12}(\kvec) =
g(\alpha_++d^{12}_\kvec \alpha_-)/\sqrt{2}$, and so on.  The result is
that $\bar H(\kvec)$ is {\it identical} to the $H_\kvec$ on the kagome
lattice in Sec. II, up to an overall additive energy $-t$.

To check the equivalence explicitly, let us solve the Hamiltonian at
one particular choice of fluxes: $\phi_+=\phi_-=0$, for example.  In
this case the spectrum is given by
\begin{eqnarray}
E &=& \pm t - g \\
E &=& \frac{g}{2}\pm\frac{1}{2}
\sqrt{4t^2+9g^2\pm 4tg\sqrt{3+q(\kvec)+\bar q(\kvec)}}
\;.
\label{eq:hex_spectrum}
\end{eqnarray}
To first order in $g/t$ the eigenvalues above
Eq.~(\ref{eq:hex_spectrum}) reduce to
\begin{eqnarray}
E&=&-t-g\\
E&=&-t+\frac{g}{2}\left(1\pm \sqrt{3+q(\kvec)+\bar q(\kvec)}\right),
\end{eqnarray}
and similarly with $t\rightarrow -t$.  This is exactly the spectrum of
a kagome tight-binding model in zero field, or type III in our
nomenclature, confirming that the kagome model is a limit of the
hexagonal model. Note that in this particular case, where time-reversal symmetry is not broken, the flat bands touch parabolically at
$\kvec=0$, as discussed before in Sec. II.  In fact, in the exact
solution in Eq.~(\ref{eq:hex_spectrum}) for all values of $g$ and $t$
the flat bands are not gapped for the same reason~\cite{B_footnote}.

\subsection{Parallel flat bands}
\label{2flatbands}

\begin{figure}
\includegraphics[angle=0,scale=0.5]{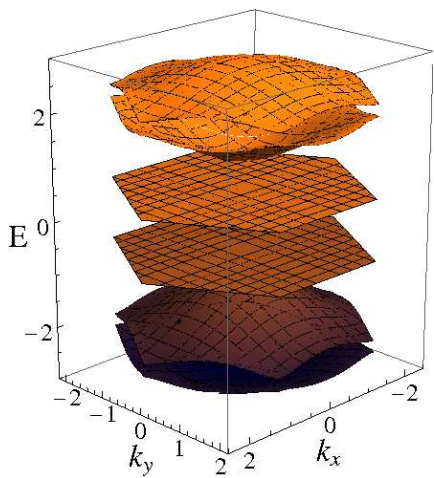}
\caption{Exact spectrum for $t=1, g=\sqrt{3}/2$ and
$\phi_\pm=3\pi/2$.}
\label{fig:parallel_bands}
\end{figure}

There is a rather interesting configuration of parameters for which
the spectrum contains two distinct flat bands. As in the kagome model,
breaking of time-reversal symmetry is required to create gapped flat
bands. We thus break time-reversal symmetry and choose the phases
$\phi_+=\phi_-=3\pi/2$. Further, we tune the interactions such that
$g=\left(\sqrt{3}/2\right) t$.  The spectrum for this Hamiltonian is
shown in Fig. \ref{fig:parallel_bands}. Notice that, remarkably,
because of the two flat bands even non-zero momenta particle-hole
excitations in this system would have macroscopic degeneracy, and
exotic heavy excitons could be formed.  

We close this section by noting that there is at least one other
hexagonal model where this parallel band structure can be induced by
breaking time-reversal symmetry. Wu and Das Sarma recently studied
ground-state properties of interacting spinless fermions in the
$p_{x,y}$-orbital bands in the two-dimensional honeycomb optical
lattice~\cite{Wu08}.  They considered a $\sigma$-bonding interaction,
which describes hopping between $p$ orbitals on neighboring sites when
the orbitals are oriented along the bond direction.  They found
gapless flat bands since their model does not break time-reversal
symmetry. However, we would like to point out that one can add to
their model a time-reversal symmetry breaking interaction, on site,
between the $p_x$ and $p_y$ orbitals of the form $gi\left(p^\dagger_x
p_y - p^\dagger_y p_x\right)$. Just as in our example, parallel flat
bands appear when $g$ is tuned to a special multiple of the hopping
strength $t$.  

\section{Summary}

Dispersionless bands can be the starting point for constructing strongly correlated electronic
states. The lack of electron kinetic energy leads to a macroscopic
degeneracy when dispersionless bands are partially filled, and
interactions become responsible for lifting the degeneracy and
selecting the many-body ground state. This situation is the case for
Landau levels, which are flat bands created by an external magnetic
field.

In this paper, we have analyzed different types of spectra that
contain flat bands in tight-binding systems in the presence of
staggered fluxes. We have seen that it is possible to separate a flat
band from the other bands by a gap when time-reversal symmetry is broken. In
these situations, the flat band can be viewed as a critical point
(with zero curvature) that separates electron-like from hole-like
bands, and we can switch between these two curvatures by changing
parameters in the Hamiltonian.  

When the gap is closed and the flat band lies between two other bands,
one obtains a spin-1 conical spectrum, extending to higher angular
momentum the spin-1/2 Dirac-like spectra in topological insulators and
in graphene. 

We have also presented examples of tight-binding systems where it is
possible to obtain more than one isolated flat band. Specifically, we
showed examples with two isolated parallel flat bands.

Although we made progress in understanding several
aspects of flat bands, two points remain open questions and 
deserve further investigation.
First, we do not have a generic proof that time-reversal symmetry must be broken to
isolate a flat band. Nonetheless, it is natural to speculate that this
is true in general, as it holds in all examples that we have found, in
addition to the well-known case of Landau levels.   
And, second, in all examples discussed here as well as in a class of tight-binding models defined on a line graph~\cite{Katsura2009}, the flat bands have zero Chern number.
Whether this is an intrinsic property of
these bands remains unclear to us. If it is possible,
however, to find examples of flat bands with 
non-zero Chern number in the absence of an external magnetic field,
this could be an interesting 
scenario for realizing strongly correlated electronic states
with topological order.  

\textit{Note added in proof.} Recently, we learned of a promising
realization of flat band systems by Koch \textit{et al.} using
circuit-QED based photon lattices.~\cite{Koch}

\section*{
Acknowledgments
         }
We thank E. Fradkin for useful discussions. This work was supported in
part by the DOE Grant No. DE-FG02-06ER46316 (C.C.).

\appendix
\section{Chern numbers for bands in the staggered flux system}
\label{sec:chern}

Haldane~\cite{Haldane1988} showed in a seminal paper that it is
possible for a system to exhibit the quantum Hall effect without
Landau levels provided the system breaks TRS. Recently, in the context
of the anomalous quantum Hall effect, Ohgushi \textit{et al}~\cite{Nagaosa}
proposed a three-band model with electrons hopping in a kagome
lattice in the presence of a background spin texture. In their model,
the spin texture opens a gap in the spectrum and gives rise to a
Berry phase such that the Chern numbers of the bands are $1,0,-1$
when TRS is broken.  In this appendix we compute explicitly the Chern
numbers in the type II spectrum of our kagome model.

Recall that when $\phi_+ + \phi_-=\pi$ and $\phi_+ - \phi_-=3\pi$ we
have a flat band with a Dirac point as shown in Fig.
\ref{fig:kagome-zero}. For this choice of fluxes, as discussed
previously, the Hamiltonian is time-reversal invariant and the system
presents no Hall response (type I). However, when the flat band is
maintained but a gap is opened, time-reversal invariance is lost and
we have the possibility of bands with non-zero Chern numbers (type
II). We parametrize the gap by a mass term $\delta$, such that the
fluxes $\phi_+=2\pi+\delta$ and $\phi_-=-\pi-\delta$.  $\delta=0$
corresponds to a Dirac cone at the center of the BZ. For reference we give the complete energy spectrum, although we
will expand around small $\vec{k}$ below: $E_\pm=\pm \sqrt{f(\kvec)}$
and $E_0=0$, where
$f(\kvec)=6-2\sum_{i,j}\cos{[\kvec\cdot(\svec_j-\svec_i)-2
\delta/3]}$, and the summation is over the cyclic permutations
$(i,j)=(1,2)$, $(2,3)$, and $(3,2)$.

The Chern number of the $n$th band is defined as the summation over
the first BZ,
\begin{eqnarray}
    C_n
    &=&\frac{-i}{2\pi}\sum_{m\neq n, \kvec \in BZ}
    \frac{\langle n\kvec|J_x|m\kvec\rangle\langle m\kvec|J_y|n\kvec\rangle
    -(J_x \leftrightarrow J_y)}
    {(E_n(\kvec)-E_m(\kvec))^2} \nonumber\\
    &=&\frac{1}{2\pi}\sum_{\kvec \in BZ}\nabla_{\kvec} \times \vec{A}_n(\kvec) \nonumber\\
    &=&\frac{1}{2\pi}\sum_{\kvec \in BZ}B_n(\kvec).
\end{eqnarray}
Here, $B_n(\kvec)$ is the field strength associated with the Berry vector
field $\vec{A}_n(\kvec)=-i\langle n\kvec|\nabla_{\kvec}|n\kvec\rangle$
and $\boldsymbol{J}=(J_x,J_y)$ is the current operator given by
$\boldsymbol{J}=\nabla_{\kvec} H$. One can see that $\sum_{n}C_n=0$ by
the antisymmetry of $C_n$ as $x$ and $y$ are interchanged. Because the
Chern numbers of the bands are topological
quantities~\cite{Thouless82,Niu}, their values can only change when a
band touching occurs. Choosing $\delta$ to be very small, an arbitrarily
small gap $ m \sim \delta $ is opened (still keeping the flat band)
and a near degeneracy appears for $\kvec \approx 0$. Around this
point, the Hamiltonian is that of a spin-$1$ system with $H_{\kvec}
\approx k_x L_x+k_y L_y+m L_z$.

To perform the summation explicitly it is convenient to define the
vector $\boldsymbol{f} \equiv (k_x,k_y,m) \equiv
|\boldsymbol{f}|(\sin{\theta}\cos{\phi},\sin{\theta}\sin{\phi},\cos{\theta})$,
which can be viewed as a magnetic field in momentum space coupled to
the spin operator (c.f., Berry~\cite{Berry}). First, we compute the
eigenvectors of $H_{\kvec}$ with the respect to the $z$ axis and then
we apply a rotation to bring the spin states to an arbitrary $(\theta,
\phi)$ direction. Let $\{|\chi_+\rangle,|\chi_0 \rangle, |\chi_-
\rangle\}$ be the eigenstates of $L_z$ with eigenvalues $1,0,-1$
respectively. The eigenvectors of $H_{\kvec}$ in a general direction
$(\theta, \phi)$ are given by
\begin{equation}
|\psi_n \rangle = e^{-i\phi L_z}e^{-i\theta L_y}|\chi_n \rangle\;,
\end{equation}
with $n=+,-,0$, which in explicit form reads
\begin{subequations}
\begin{equation}
|\psi_+ \rangle=
|
\begin{array}{ccc}
 e^{-i\phi}\left( \frac{1+\cos{\theta}}{2} \right), & \frac{\sin{\theta}}{\sqrt{2}}, & e^{i\phi}\left(\frac{1-\cos{\theta}}{2}\right)
\end{array}
\rangle
\end{equation}
\begin{equation}
|\psi_0 \rangle=
|
\begin{array}{ccc}
 -e^{-i\phi}\frac{\sin{\theta}}{\sqrt{2}}, & \cos{\theta}, & e^{i\phi}\frac{\sin{\theta}}{\sqrt{2}}
\end{array}
\rangle
\end{equation}
\begin{equation}
|\psi_- \rangle=
|
\begin{array}{ccc}
 e^{-i\phi}\left(\frac{1-\cos{\theta}}{2}\right), & -\frac{\sin{\theta}}{\sqrt{2}}, & e^{i\phi}\left(\frac{1+\cos{\theta}}{2}\right)
\end{array}
\rangle
\;,
\end{equation}
\end{subequations}
and $H_{\kvec}|\psi_{\pm} \rangle=\pm \sqrt{|\kvec|^2 +
m^2}|\psi_{\pm} \rangle$ and $H_{\kvec}|\psi_{0} \rangle=0$. A
straightforward calculation of the field strengths gives us
\begin{equation}
B_{\pm}(\kvec) = \pm \frac{m}{(m^2+|\kvec|^2)^{3/2}}~~,~~ B_0(\kvec) = 0
\;.
\end{equation}
The contributions of these fluxes to the Chern numbers are found to be
$\pm \text{sgn}(m)$ and zero.  We have also confirmed this result numerically
over the entire BZ without linearizing around the $\Gamma$ point. As
we cross the gap, $\delta$ (equivalently, $m$) changes sign and the
Chern numbers of the upper and lower bands change sign as well, while
the Chern number of the flat band remains zero. Because the
topologolical nature of the Chern numbers, their values will remain
unaltered until a new band touching occurs, which will happen for
$\delta=\pm \pi$, when the Dirac point moves to one of the corners of
the BZ.



\begin{thebibliography}{99}
\bibitem{Straley_footnote} In models for amorphous semiconductors, the Hamiltonian can be written as $H = \hat{A}+\hat{T}$, where $\hat{A}$
accounts for the intra-cluster hopping and satisfies $\hat{A}^2 =
\hat{A}$, {\it i.e.}, it is a projection operator, and $\hat{T}$ is the
operator responsible for the hopping between nearest-neighbor clusters
and satisfies $\hat{T}^2=I$. These properties of $\hat{A}$ and
$\hat{T}$ endow the spectrum with dispersionless
bands.
\bibitem{B_footnote} We speculate that the flat band arises from the property of the on-site interaction in the hexagonal Hamiltonian, $B^2 = B^\dagger$.  This would be a generalization of Straley's result for amorphous semiconductors, where the intra-cluster hopping satisfies $\hat{A}^2 = \hat{A}$, although we do not have an exact proof.
\end{thebibliography}
\end{document}